\documentclass[]{aa}
\usepackage{natbib}
\usepackage{aas_macros}
\bibpunct{(}{)}{;}{a}{}{,}
\usepackage{graphicx}
\usepackage{amssymb}
\usepackage{color}

\begin{document}

\title{Which Massive stars are Gamma-Ray Burst Progenitors?}

\author{Jelena Petrovic\inst{1,2}
       \and
        Norbert Langer\inst{1}
       \and Sung-Chul Yoon\inst{1,3}
       \and Alexander Heger\inst{4,5}
        }

\authorrunning{Petrovic et al.}
\titlerunning{Gamma-Ray Burst Progenitors}

\institute{Sterrenkundig Instituut, Utrecht University,
           Princetonplein 5,
           3584 CC Utrecht,
           The~Netherlands
	   \and
	   Astronomical Institute,
	   Radboud Universiteit Nijmegen, 
	   Toernooiveld 1, 6525 ED,
	   Nijmegen, The~Netherlands
           \and
	   Astronomical Institute "Anton Pannekoek", 
	   Amsterdam University
           Kruislaan 403, 1098 SJ, 
	   Amsterdam, The~Netherlands
	   \and
           Theoretical Astrophysics Group, 
           T-6, MS B227,  
           Los Alamos National Laboratory, 
           Los Alamos, NM 87545, U.S.A.
           \and
	   Enrico Fermi Institute, 
           The University of Chicago,
	   5640 S. Ellis Ave, 
           Chicago, IL 60637, U.S.A.}

\date{Received; accepted} 
\offprints{J.Petrovic,\\ \email{petrovic@astro.ru.nl}}
\abstract{The collapsar model for gamma-ray bursts requires three
essential ingredients: a massive core, removal of the hydrogen
envelope, and enough angular momentum in the core.  We study current
massive star evolution models of solar metallicity to determine which
massive star physics is capable of producing these ingredients. In
particular, we investigate the role of hydrodynamic and magnetic
internal angular momentum transport and binary mass and angular
momentum transfer.  We follow the evolution of rotating single stars
and of binary systems that include rotational processes for both
stars.  Neglecting magnetic fields, we show that the cores of massive
single stars can maintain a high specific angular momentum
($j$$\sim$10$^{17}\rm~cm^2~s^{-1}$) when evolved with the assumption
that mean molecular weight gradients suppress rotational mixing
processes.  In binary systems that undergo mass transfer during core
hydrogen burning the mass receiving star accretes large amounts of
high angular momentum material, leading to a spin-up of the core.  We
find, however, that this merely compensates for the tidal angular
momentum loss due to spin-orbit coupling, which leads to synchronous
rotation before the mass transfer event.  Therefore the resulting
cores do not rotate faster than in single stars.  We show that some
accreting stars become Wolf-Rayet stars at core helium exhaustion and
form CO-cores that are massive enough to form a black hole.  We also
present models that include magnetic fields generated by differential
rotation and we consider the internal angular momentum
transport by magnetic torques.  Though magnetic single star models are
known to develop rather slowly rotating cores 
with specific angular momenta at the end of the evolution close to
those in observed young pulsars ($j$$\sim$10$^{14}\rm~cm^2~s^{-1}$),
we investigate the capability of magnetic torques to efficiently pump
angular momentum into the cores of accreting stars.  Despite our
finding that this mechanism works, the magnetic coupling of core and
envelope after the accreting star ends core hydrogen burning leads to
slower rotation ($j$$\sim$10$^{{15}-{16}}\rm~cm^2~s^{-1}$) than in the
non-magnetic case.  We conclude that our binary models without
magnetic fields can reproduce stellar cores with a high enough
specific angular momentum ($j$$\ge$3$\cdot$10$^{16}\rm~cm^2~s^{-1}$)
to produce a collapsar and a GRB.  If magnetic torques are included,
however, GRBs at near solar metallicity need to be produced in rather
exotic binary channels, or current dynamo model overestimates
the magnetic torques.  But then the problem is that significant angular
momentum loss from the iron core either during core collapse or from
the proto-neutron star would be required.  
 }

\maketitle
\section{Introduction}\label{intro} 

From studies of host galaxies of gamma-ray bursts it was concluded
that they occur in or close to star forming regions.  Several of them
are associated with a very energetic variety of Type~Ic supernova with
broad lines \citep["hypernova",][]{2002ApJ...572L..51P,2003Natur.423..847H},
thought to display the explosion of a
massive Wolf-Rayet star.  And finally, their afterglows show
signatures of the shaping of the circumstellar medium by a massive
progenitor star \citep{marlepaper}.  The most widely used model for
GRB production in the context of black hole formation in a massive
single star is the so called collapsar model
\citep{1993AAS...182.5505W}.

A collapsar is a massive \citep[$M\,$$\gtrsim$$\,35$-$40\,\mathrm 
{M}_{\odot}$,][]{1999ApJ...522..413F} rotating star whose core collapses
to form a black hole \citep{1993ApJ...405..273W, 1999ApJ...524..262M}.
If the collapsing core has enough angular momentum
\citep[$j$$\ge$3$\cdot$10$^{16}\rm~cm^2~s^{-1}$,][]{1999ApJ...524..262M} an accretion disk is formed around
the black hole.  The accretion of the rest of the core at accretion
rates up to 0.1$\,\mathrm{M}_{\odot}s^{-1}$ by the newly-formed black hole is
thought to be capable of producing a collimated highly relativistic
outflow.  This releases large amounts of energy ($\sim$10$^{51}\rm~
erg~s^{-1}$) some of which is deposited in the low density rotation
axis of the star.  In case the star has no hydrogen envelope, i.e.,
has a light crossing time which is less or comparable to the duration
of the central accretion (about 10s), a GRB accompanied by a Type Ib/c
supernova may be produced.  The collapsar models for gamma-ray bursts
thus need three essential ingredients: a massive core, loss of the
hydrogen envelope, and sufficient angular momentum to form an
accretion disk.

\citet{2000NewAR..44..297H} have calculated models of a 25$\,\mathrm
M_{\odot}$ star that could form a black hole by fallback (SN explosion
occurs while the core forms a neutron star, but so much matter fails
to escape and falls back onto the neutron star that it turns into a
black hole).  This star ends its life as a red supergiant with an iron
core of 1.9$\,\mathrm{M}_{\odot}$, a helium core of 8.06$\,\mathrm{M}_{\odot}$, and
a low density envelope of 6.57$\,\mathrm{M}_{\odot}$.  They found that this
star has sufficient angular momentum to form an accretion disk around
the black hole which may lead to an asymmetric, jet-driven supernovae,
or in the case that the star lost its hydrogen envelope, even a GRB
can result.  The rotating pre-supernova models of
\citet{2000ApJ...528..368H} predicted, for the mass range 10...20$\,\mathrm
M_{\odot}$, an iron core angular momentum barely sufficient for the
GRB production model through collapsar, with a trend of decreasing
final specific angular momentum for larger initial mass.
\citet{2000ApJ...528..368H} showed also that specific angular momentum
of the stellar core depends significantly on the inhibiting effect of
the mean molecular weight gradient on rotational mixing processes.

Only if composition is a efficient barrier for rotational mixing and
transport of angular momentum, it prevents the core from losing most
of its angular momentum during the evolution.  Finally, the rotating
pre-supernova models of \citep{2004A&A...425..649H} predict about twice as
much final core angular momentum as the models of
\citet{2000ApJ...528..368H}.

These calculations left out the influence of a magnetic field which
can significantly alter the angular momentum transport processes in
the star, for example slowing down the helium core of the star during
the red supergiant phase \citep{1998Natur.393..139S}.  Single star pre-supernova
models including angular momentum transport by magnetic torques, using
the improved dynamo model of \citet{2002A&A...381..923S}, have been
produced by \citet{2004IAUS..215b} and \citet{2004astro.ph..9422H}, with a clear result: although these models predict
neutron star spins in the range displayed by young pulsars, the amount
of angular momentum in their cores is one to two orders of magnitude
less than what is required by the collapsar model of GRB production.
A possible conclusion therefore might be that, if the dynamo model of
\citet{2002A&A...381..923S} is qualitatively right, massive single
stars can not form GRBs within the collapsar model.

Only one in one hundred collapsing massive stars needs to produce,
however, a GRB in order to explain their frequency. It seems therefore
legitimate to explore collapsars in the frame of massive close binary
evolution.  \citet{wellsteinphd} and \citet{2004AAinprep} showed that
during the mass transfer phase in a binary system, the secondary
(accreting) star can spin up to close to critical rotation, i.e.,
surface layers of this star can gain large amounts of angular
momentum.  This angular momentum can be transported inward and
increase the rotation rate of the stellar core.  In this case, one may
expect that the effects of magnetic fields would be helpful in
spinning up the core, and that faster rotating cores might be produced
than in the non-magnetic models.

In this paper, we investigate four different types of rotating massive
star models: single and binary models, both with and without magnetic
fields included.  In Sect.~\ref{comp} we briefly explain our numerical
methods and physical assumptions, and in Sect.~\ref{nonmagnetic} we
explore non-magnetic single and binary stars.  In Sect.~\ref{magnetic}
we present our magnetic models, first single stars then binary evolution
models, and in Sect.~\ref{concl} we summarize our conclusions.

\section{Computational method}\label{comp}

We compute detailed evolutionary rotating models of non-magnetic
20$\,\mathrm{M}_{\odot}$ and 42$\,\mathrm{M}_{\odot}$ single stars, and of a
42$\,\mathrm{M}_{\odot}$ single star with magnetic field included.  Then we
calculate the evolution of a non-magnetic and a magnetic binary system
with rotating components of $M_{\rm 1,in}$=56$\,\mathrm{M}_{\odot}$ and
$M_{\rm 2,in}$=33$\,\mathrm{M}_{\odot}$ and an orbital period $p_{\rm in}$=6
days.  We also compute models of a rotating 33$\,\mathrm{M}_{\odot}$ star
that accretes matter at the end of its main sequence evolution, for
the non-magnetic and magnetic cases.

We use the evolutionary code developed by \citet{1998PhDT........29B}
on the basis of an implicit hydrodynamic stellar evolution code for
single stars \citep{1991A&A...252..669L,1998A&A...329..551L}.

The treatment of convection and semiconvection have been described in
\citet{1991A&A...252..669L} and \citet{1995A&A...297..483B}.  Changes
in chemical composition are computed using a nuclear network including
pp chains, the CNO-cycle, and the major helium, carbon, neon and
oxygen burning reactions.  More details are given in
\citet{1999A&A...350..148W} and \citet{2001A&A...369..939W}.  For all
models, a metallicity $Z$=0.02 is adopted and we use the OPAL
opacities \citep{1996ApJ...464..943I}.  The abundance ratios of the
isotopes are chosen to have the solar meteoritic
abundance ratios according to
\citet{1993ist..proc..410N}. The change of the orbital
period due to mass transfer and stellar wind mass loss is computed
according to \citet{1992ApJ...391..246P}, with a specific angular
momentum of the stellar wind material calculated according to
\citet{1993ApJ...410..719B}.  The influence of the centrifugal force
in the rotating models is implemented according to
\citet{1970stro.coll...20K}.

Stellar wind mass loss for O stars is calculated according to
\citet{1989A&A...219..205K}.  For hydrogen poor stars ($X_{\rm
s}<0.4$) we used a relation based on the empirical mass loss rates of
Wolf-Rayet stars derived by \citet{1995A&A...299..151H}:
\begin{equation}\label{maslos}
{log({\dot M_{\rm WR}}/{\,\mathrm{M}_{\odot}yr^{\rm -1}})}=-11.95+1.5logL/L_{\odot}-2.85X_{\rm s}.
\end{equation}
Since \citet{1998A&A...335.1003H} suggested that these mass loss rates
may be overestimated, we calculated evolutionary models in which the
mass loss rate given by Eq.~\ref{maslos} is multiplied by $1/3$.

Also, enhanced mass loss due to rotation is included:
\begin{equation}
{\dot M}/{\dot M (v_{\rm rot}=0)}={1}/{(1-{\Omega})^{\rm \xi}},
\end{equation}
where $\xi$=0.43, $\Omega$=$v_{\rm rot}/v_{\rm crit}$ and $v_{\rm
crit}^{\rm 2}$=${GM(1-\Gamma)}/R$ with $\Gamma$=$L/L_{\rm
Edd}$=${\kappa}L/(4{\pi}cGM)$ is the Eddington factor, $G$ is
gravitational constant, $M$ is mass, $R$ radius, $\kappa$ opacity,
$v_{\rm rot}$ rotating velocity and $v_{\rm crit}$ critical rotational
velocity \citep{1998A&A...329..551L}.

The transport of angular momentum in our code is formulated as a
diffusive process:
\begin{eqnarray}
\left (\frac{\partial \omega}{\partial t} \right )_m={\frac{1}{i}}
\left (\frac{\partial}{\partial m} \right )_t \left [ \left (4\pi r^2
\rho \right )^2i \nu \left (\frac{\partial \omega}{\partial m} \right
)_t \right ] \nonumber\\ -{\frac{2w}{r}} \left (\frac{\partial
r}{\partial t} \right )_m{\frac{1}{2}}{\frac{dlni}{dlnr}},
\end{eqnarray}
where $\nu$ is the turbulent viscosity and $i$ is the specific angular 
momentum of a shell at mass coordinate $m$. Factor $(1/2)(dlni/dlnr)$
vanishes if the gyration constant $k=i/r^2$ does not depend on $r$,
i.e. if all shells are moving homologously.

A parameter $f_{\mu}$=0.05 is adopted for sensitivity of the
rotationally induced mixing processes to the $\mu$-gradient
\citep{2000ApJ...528..368H}.
As shown by \citet{1997A&A...321..465M}, $\mu$-gradients can
efficiently suppress rotationally induced transport processes. The
strenght of this inhibiting effect is descibed by the parameter
$f_{\mu}$=$0..1$ \citep{1989ApJ...338..424P}.  A value of
$f_{\mu}$=$0.05$ reproduces the best observations of the enrichement in
the surface of the main sequence stars with products of CNO process.

At the surface of the star, the angular momentum contained in the
layers which are lost due to stellar wind gets removed from the star:
\begin{equation}
\dot J=\dot M j_{\mathrm{spec}}
\end{equation}
where $j_{\mathrm{spec}}$ is the average angular momentum at the
surface of the star and $\dot M$ the stellar wind mass loss rate.  The
turbulent viscosity, $\nu$, is determined as the sum of the convective
and semiconvective diffusion coefficients
and those from rotationally induced instabilities
(dynamical shear, Solberg-H{\o}iland, secular shear, 
Goldreich-Schubert-Fricke instability and Eddington-Sweet
circulation).  Rotationally induced mixing processes and angular
momentum transport through the stellar interior are described in
detail by \citet{2000ApJ...528..368H}.

The code calculates the simultaneous evolution of two stellar
components of a binary system and computes mass transfer within the
Roche approximation \citep{1978dcbs.conf.....K}.  Mass loss from the
Roche lobe filling component through the first Lagrangian point is
given by \citet{1988A&A...202...93R} as:
\begin{equation}
{\dot M}={\dot M_{\rm 0}}\exp(R-R_{\rm l})/{H_{\rm p}}
\end{equation}
with ${\dot M_{\rm 0}}$=${{\rho}v_{\rm s}Q}/{\sqrt e}$, where $e$ is
the base of the natural logarithm, $H_{\rm p}$ is the photospheric
pressure scale height, $\rho$ is the density, $v_{\rm s}$ the velocity
of sound, and $Q$ the effective cross-section of the stream through the
first Lagrangian point according to \citet{1983A&A...121...29M}.  The
time scales for synchronization and circularization of the binary
orbit as well as spin-orbit coupling are given by
\citet{1977A&A....57..383Z}.  The specific angular momentum of the
accreted matter is determined by integrating the equation of motion of
a test particle in the Roche potential \citep[ accretion stream
impacts directly on the secondary star]{wellsteinphd}.

We calculate the evolution of binary systems in detail until Case AB
mass transfer starts.  Then we remove the hydrogen rich envelope from
the primary, until only 5\% of the hydrogen is left in the
envelope. This is the point where we assume that the primary shrinks
and Case AB stops.  On the other side, we calculate the secondary star
assuming accretion with a mass transfer of 10$^{-4} \,\mathrm
{M}_{\odot} \rm yr^{-1}$.  We calculate the Kelvin-Helmholtz time scale of
the primary on the beginning of Case AB mass transfer:
\begin{equation}
{t_{\rm KH}=2\cdot10^{\rm 7}{M_{\rm 1}}^{\rm 2}/(L_{\rm 1}R_{\rm l1})\rm yr}, 
\end{equation}
where $M_{\rm 1}$, $L_{\rm 1}$ and $R_{\rm l1}$ are mass, luminosity
and Roche radius (in Solar units) of the primary star at the onset of
Case AB mass transfer.  Mass transfer rate is:
\begin{equation}\label{mtr}
{\dot M_{\rm tr}=(M_{\rm 1}-M_{\rm He})/{t_{\rm KH}}}
\end{equation}
where $M_{\rm He}$ is the mass of the helium core with $\sim$5\% of
hydrogen on the surface, i.e. the initial WR mass.  We calculate the
orbit change assuming constant mass transfer rate calculated by
Eq.~\ref{mtr} and an average accretion efficiency ($\beta$) which is
the same as for the fast phase of Case A, since Case AB also happens
on the thermal time scale.  Matter that has not been accreted on the
secondary leaves the system with a specific angular momentum which
corresponds to the secondary orbital angular momentum
\citep{2001MNRAS.321..327K}.  Stellar wind mass loss is neglected.

We follow the evolution of the binary system, again in detail, until
the primary finishes carbon core burning.  We assume that after this
the system is disrupted by the SN explosion of the primary and we
model the secondary further as a single star. Stellar wind mass loss
for red supergiants is given by \citet{1990A&A...231..134N}.

Magnetic fields generated by differential rotation are included
according to \citet{2002A&A...381..923S}.  The rate at which the
field is amplified is determined by the differential rotation.
Differentional rotation rate (rotation gradient) is a function of a
radial coordinate only ('shellular rotation') and is given by:
\begin{equation}
q=\frac{\partial\ln\Omega_\star}{\partial\ln r}=\frac{r\partial_r \Omega_\star}{\Omega_\star}
\end{equation}
The dynamo process requires a minimum rotation gradient $q_{min}$ to
operate.

The effective radial viscosity produced by the magnetic field is:
\begin{equation}
\nu_{re}=\frac{\nu_{e0}\nu_{e1}}{\nu_{e0}+\nu_{e1}}f(q),
\end{equation}
where
\begin{equation}
\nu_{e0}=r^2\Omega_\star q^2 \left (\frac{\Omega_\star}{N_{\mu}} \right )^4,
\end{equation}

\begin{equation}
\nu_{e1}=r^2\Omega_\star max \left [\left (\frac{\Omega_\star}{N_T} \right )^{1/2} \left (\frac{\kappa}{r^2N_T}
\right )^{1/2},q^2\left (\frac{\Omega_\star}{N_T} \right )^4 \right ],
\end{equation}
$r$ is the radial coordinate, $\Omega_\star$ is the rotation rate of
the star, is $N_{\mu}$ the compositional buoyancy frequency, $N_T$ is the
thermal part of buoyancy frequency, $q$ is the rotational gradient and
$q_{\rm min}$ is minimum rotational gradient necessary for the dynamo
to operate:
\begin{equation}
f(q)=1-q_{\rm min}/q,  (q>q_{\rm min}),
\end{equation}
and
\begin{equation}
f(q)=0, (q\le q_{\rm min}).
\end{equation}  
The factor $f(q)$ causes the stress to vanish smoothly as the gradient
of the rotation rate approaches the minimum value required for dynamo
action.

Fluid motions involved in the dynamo process also imply a certain
amount of mixing.  The effective diffusivity is given as:
\begin{equation}
D_e=\frac{D_{e0}D_{e1}}{D_{e0}+D_{e1}}f(q),
\end{equation}  
where $f(q)$ is defined by Eq.(11) and
\begin{equation}
D_{e0}=r^2\Omega_\star q^4 \left (\frac{\Omega_\star}{N_{\mu}} \right )^6,
\end{equation}

\begin{equation}
D_{e1}=r^2\Omega_\star max \left [ \left (\frac{\Omega_\star}{N_T} \right )^{3/4} \left (\frac{\kappa}{r^2N_T}
\right )^{3/4},q^2 \left (\frac{\Omega_\star}{N_T} \right )^6 \right]. 
\end{equation}

\section{Models without magnetic field}\label{nonmagnetic}

\subsection{Single stars}\label{single}
It was shown by \citet{1997A&A...321..465M} and
\citet{2000ApJ...528..368H} that transport of angular momentum in
stellar interiors depends strongly on the efficiency of the inhibition
of the rotationally induced mixing processes by $\mu$-gradient.  When
rotationally induced mixing processes are not inhibited by
$\mu$-gradient, stellar rotation remains close to rigid rotation. The
star loses matter from the surface which carries away angular
momentum.  Layers beneath the lost ones expand, and due to local
angular momentum conservation, spin down.  The star reestablishes
rigid rotation by transporting angular momentum from the core to the
surface.  When the rotationally induced mixing processes are inhibited
by the $\mu$-gradient, angular momentum cannot be efficiently
transported between the core and the envelope.  The result is
differential rotation between these two stellar regions.  In this
paper we present models which include the inhibiting effect of
$\mu$-gradient on rotationally induced mixing processes.

\begin{figure}
  \centering
   \includegraphics[width=\columnwidth]{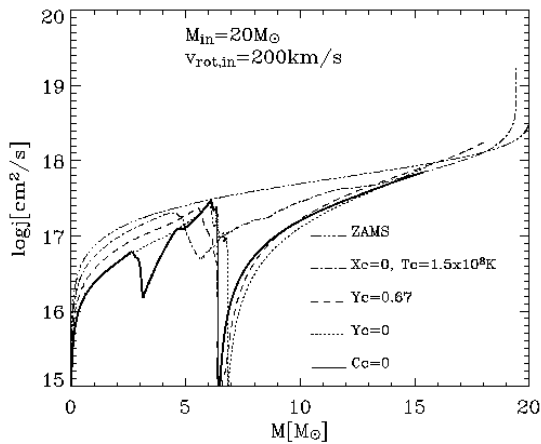}
  \caption[]{\label{core20}Specific angular momentum profiles of a
  20$\,\mathrm{M}_{\odot}$ single star on the hydrogen ZAMS (three
  dots-dashed line), when helium ignites in the core, (dot-dashed
  line), when the central helium abundance is 67\% (dashed-line), in
  the moment of core helium exhaustion (dotted line) and at core
  carbon exhaustion (solid line.)}
\end{figure}

We calculate the evolution of a 20$\,\mathrm{M}_{\odot}$ star until the end
of core carbon burning.  This star is a rigidly rotating star on the
ZAMS (Fig.~\ref{core20}, three dots-dashed line) with an initial
surface velocity of $v_{\rm surf}$=200$\rm~km~s^{-1}$, which is a
typical value for these stars 
\citep{1982PASP...94..271F}. During its hydrogen core burning phase,
this star loses mass due to a stellar winds ($\sim$10$^{-8}\,\mathrm
{M}_{\odot} \rm~yr^{-1}$). Matter lost from the stellar surface carries away
angular momentum, the surface layers spin down, but there is no
efficient transport of angular momentum between the core and the
envelope, so the core does not spin down significantly. At the time of
helium ignition (Fig.~\ref{core20},dot-dashed line), the specific
angular momentum of the core is $j$$\sim$10$^{17} \rm~cm^2~s^{-1}$ at
3$\,\mathrm{M}_{\odot}$ (further in the paper, specific angular momentum
values are always given at 3$\,\mathrm{M}_{\odot}$, since that is the mass of
the final core before supernova explosion).  After core hydrogen
exhaustion, the star evolves into a red supergiant, its core contracts
while the envelope expands.  This leads to a spin-up of the core and a
spin-down of the envelope.  The envelope is convective and rotating
almost rigidly with a very low rotational velocity ($\sim$0.1$\rm~
km~s^{-1}$). The core is rigidly rotating with a maximum rotational
velocity of $\sim$130$\rm~km~s^{-1}$.  The core and the envelope are
separated by layers that have a large $\mu$-gradient.  This suppresses
rotationally induced mixing and angular momentum is not efficiently
transported, so that the core is not slowed down by the slow rotation
of the envelope.  At the end of helium core burning the specific
angular momentum of the core is $j$$\sim$5$\cdot$10$^{16} \rm~cm^2~s^{-1}$ (Fig.~\ref{core20}, dotted line).  During further evolution,
i.e., core carbon burning, the stellar core does not lose any
significant amounts of angular momentum.

\begin{figure}
  \centering
   \includegraphics[width=\columnwidth]{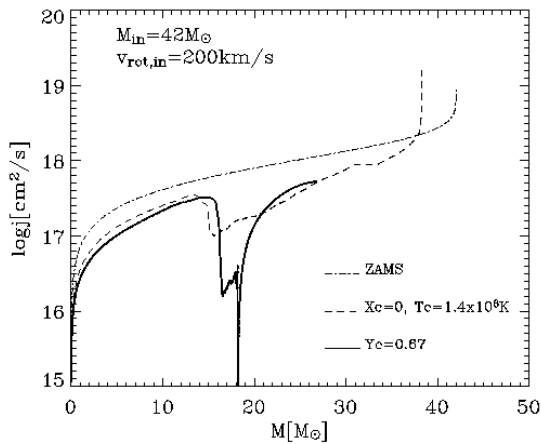}
  \caption[]{\label{core42}Specific angular momentum profiles of a
  42$\,\mathrm{M}_{\odot}$ single star on the hydrogen ZAMS (dot-dashed
  line), when helium ignites in the center (dotted line), and when the
  central helium abundance is 67\% (solid line).}
\end{figure}

To investigate the behaviour of higher mass stars, we modeled the
evolution of a 42$\,\mathrm{M}_{\odot}$ star with the same initial surface
rotational velocity $v_{\rm surf}$=200$\rm~km~s^{-1}$. The star is a
rigidly rotating star on the ZAMS with the specific angular momentum
profile shown on Fig.~\ref{core42} (dot-dashed line).  As we already
explained for the previous example, the star loses mass due to a
stellar wind ($\sim$10$^{-7}\,\mathrm{M}_{\odot} \rm~yr^{-1}$).  Since the stellar
wind mass loss rate is one order of magnitude higher than for a 20$\,\mathrm
{M}_{\odot}$ star, the 42$\,\mathrm{M}_{\odot}$ star loses more matter and
angular momentum during the core hydrogen burning phase.  When helium
ignites in the core, the specific angular momentum at 3$\,\mathrm{M}_{\odot}$
is $\sim$8$\cdot$10$^{16}\rm~cm^2~s^{-1}$.  The star becomes a
$\sim$38$\,\mathrm{M}_{\odot}$ red supergiant with central helium burning in
a convective core of $\sim$12$\,\mathrm{M}_{\odot}$ and a convective envelope
of $\sim$20$\,\mathrm{M}_{\odot}$. Due to stellar wind mass loss of
$\sim$10$^{-4} \,\mathrm{M}_{\odot} \rm~yr^{-1}$, the mass of the envelope rapidly
decreases and reaches $\sim$10$\,\mathrm{M}_{\odot}$ at the end of our
calculations.

We follow the evolution of this star until it burned $\sim$30\% of the
helium in its core.  The specific angular momentum of the core (at
3$\,\mathrm{M}_{\odot}$) at that moment is $\sim$6.5$\cdot$10$^{16}\rm~cm^2~s^{-1}$. 
If we assume that the angular momentum of the core decreases
further during helium core burning with the same rate, the specific
angular momentum of the core at the moment of helium exhaustion would
be $\sim$3.5$\cdot$10$^{16}\rm~cm^2~s^{-1}$.  We see in
Fig.~\ref{core20} that there is no angular momentum loss from the core
during core carbon burning, so we can conclude that this star might
produce a collapsar and in case that the hydrogen envelope is lost
during the red the supergiant phase, a gamma-ray burst can be the
result.

\begin{figure}
  \centering
   \includegraphics[width=\columnwidth]{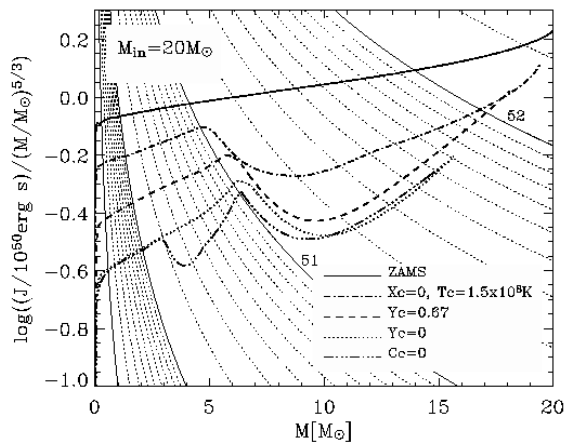}
  \caption[]{\label{jt20}Logarithm of the integrated angular momentum,
  $J(m)$=$\int_{0}^{M}j(m^{\prime})dm^{\prime}$ divided by $m^{5/3}$,
  as a function of the mass coordinate, $m$, for a 20$\,\mathrm{M}_{\odot}$
  star for the same evolutionary stages as shown in
  Fig.~\ref{core20}. The thin lines give a logarithmic scale of levels
  of constant $J$ labeled with $\rm log(J/\rm (erg~s))$.}
\end{figure}

\begin{figure}
  \centering
   \includegraphics[width=\columnwidth]{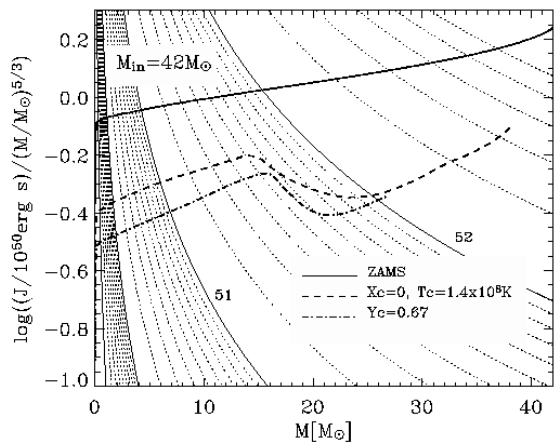}
  \caption[]{\label{jt42}Logarithm of the integrated angular momentum
  $J(m)$=$\int_{0}^{M}j(m^{\prime})dm^{\prime}$ divided by $m^{5/3}$,
  as a function of the mass coordinate $m$ for a 42$\,\mathrm{M}_{\odot}$
  star for the same evolutionary stages as shown on
  Fig.~\ref{core42}. The thin lines give a logarithmic scale of levels
  of constant $J$ labeled with $\rm log(J/\rm (erg~s))$.}
\end{figure}

For a rigidly rotating body of constant density $\rho_0$, the total
angular momentum, $J(m)$, enclosed below the mass coordinate, $m$, is
$J(m)$=$\int_{0}^{M}j(m^{\prime})dm$. In Fig.~\ref{jt20} and
Fig.~\ref{jt42} we plot the logarithm of $J(m)/m^{5/3}$ as a
function of mass and a grid of lines of $J$=$const$.  If angular
momentum is transported throughout the surface defined by a mass
coordinate $m$, $J(m)$ drops. If no such transport occurs, $J(m)$
remains constant.  If a line of constant $J$ is followed from one
evolutionary stage to another, it can be seen to what mass coordinate
the angular momentum has been transported in the star during the time
between the two evolutionary stages.

\subsection{Binary systems}\label{binary}

A star evolving in a binary system and accreting matter from the
companion, increases its surface angular momentum. If this angular
momentum can be transported efficiently through the stellar interior,
the star may evolve into a red supergiant that has a rapidly spinning
core with sufficient specific angular momentum to produce a collapsar.

To check if accretion can add enough angular momentum to the core, we
modeled the evolution of a rotating binary system with initial masses
of $M_{\rm 1,in}$=56$\,\mathrm{M}_{\odot}$ and $M_{\rm 2,in}$=33$\,\mathrm
{M}_{\odot}$ and an initial orbital period of $p_{\rm in}$=6 days.  The
binary system quickly synchronizes during the main sequence evolution.
Due to this synchronization, both stars lose angular momentum and
their initial surface rotational velocities are 92$\rm~km~s^{-1}$ for
the primary and 64$\rm~km~s^{-1}$ for the secondary which is much
slower than the typical values for single stars of these masses
\citep{2000ApJ...528..368H}. This means that stars in binary systems
lose a significant amount of angular momentum due to synchronization.
The angular momentum loss increases with the initial orbital period.

\begin{figure}
  \centering
   \includegraphics[width=\columnwidth]{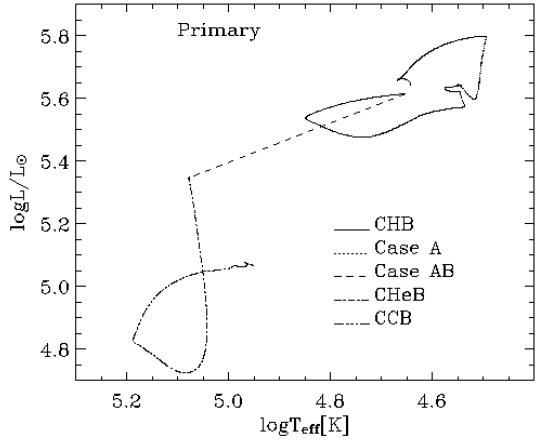}
  \caption[]{\label{prim}Evolutionary track of the primary in HR
  diagram.  Solid line: core hydrogen burning phase before and after
  Case A mass transfer. Dotted line: Case A mass transfer.  Dashed
  line: connection between last calculated model at the onset of Case
  AB and first calculated model when the primary is a hydrogen-free WR
  star. Dash-dotted line: core helium burning phase. Dot-dashed line:
  core carbon burning phase. }
\end{figure}

\begin{figure}
  \centering
   \includegraphics[width=\columnwidth]{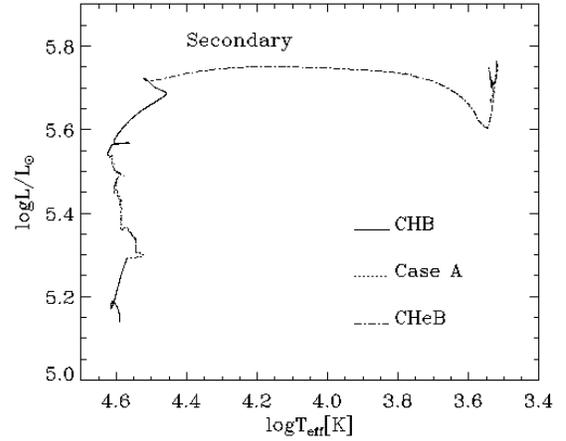}
  \caption[]{\label{secon}Evolutionary track of the secondary in the
  HR diagram.  Solid line: core hydrogen burning phase before and
  after Case A and Case AB mass transfer.  Dotted line: Case A mass
  transfer. Dash-dotted line: core helium burning phase.}
\end{figure}

Evolutionary tracks of the primary and the secondary star are given in
Fig.~\ref{prim} and Fig.~\ref{secon}.  The primary is initially the
more massive star; it evolves faster and fills its Roche lobe during
hydrogen core burning.  The binary system enters Case~A of mass
transfer (dotted line, Fig.~\ref{prim}).  The primary loses matter
with a high mass transfer rate ($\dot M_{\rm tr}^{\rm
max}$$\sim$3.2$\cdot$10$^{-3} \,\mathrm{M}_{\odot} \rm~yr^{\rm -1}$) and decreases
in luminosity.  During the fast phase of Case A mass transfer, the
primary loses $\sim$19$\,\mathrm{M}_{\odot}$ and the secondary accretes only
about $15$\% of that matter due to the stellar wind mass loss caused
by rotation close to break-up \citep{2004AAinprep}.  After the fast
process of Case A mass transfer, the primary continues to expand on a
nuclear time scale and transfer mass onto the secondary star with
mass transfer rate of $\sim$10$^{-6}\,\mathrm{M}_{\odot} \rm~yr^{-1}$ (slow phase
of Case A).  At the end of core hydrogen burning the primary contracts
and thus Roche lobe overflow (RLOF) stops. When the primary starts
hydrogen shell burning ($M$=16.6$\,\mathrm{M}_{\odot}$) it expands again,
fills its Roche lobe and Case AB mass transfer starts. We assume that
this mass transfer stops when the WR star has 5\% of hydrogen on its
surface ($M_{\rm WR}(X_{\rm s}$=0.05)=14.8$\,\mathrm{M}_{\odot}$).  More
details about the evolution of this system up to WR+O phase can be
seen in \citet{2004AAinprep}.  As explained in Sect.~\ref{comp}, we
continue following the evolution of the binary system with a
hydrogen-free WR star. The dashed line on Fig.~\ref{prim} connects the
last calculated model at the onset of Case AB and the first calculated
model when the primary is a hydrogen-free WR star.  The initial
hydrogen-free WR star is $M_{\rm WR}$=12.8$\,\mathrm{M}_{\odot}$ with an
effective temperature of $\sim$1.2$\cdot$10$^5 \rm~K$, and it loses
$\sim$8$\,\mathrm{M}_{\odot}$ due to WR mass loss during core helium
burning. Due to this mass loss, the luminosity of the primary
decreases (dash-dotted line Fig.~\ref{prim}).  The primary ends carbon
core burning (dot-dashed line Fig.~\ref{prim}) as $\sim$4.5 solar mass
star.  The orbital period of the binary system is $\sim$13.5 days when
the primary explodes in a supernova.

The secondary begins its evolution as a 33$\,\mathrm{M}_{\odot}$ core
hydrogen burning star.  It accretes $\sim$3$\,\mathrm{M}_{\odot}$ during the
fast phase and $\sim$4$\,\mathrm{M}_{\odot}$ during the slow phase of Case A
mass transfer.  Due to this mass gain, the luminosity of the secondary
increases (Fig.~\ref{secon}). As we already mentioned, the primary
loses $\sim$1.8$\,\mathrm{M}_{\odot}$ during Case AB.  We assume the same
average accretion efficiency during this mass transfer as during the
fast phase of Case A mass transfer ($\beta$=0.15), since they both
take place on the thermal time scale.  This means that the secondary
accretes $\sim$0.25$\,\mathrm{M}_{\odot}$ and after Case AB mass transfer, it
is still a core hydrogen burning star, more massive than at the
beginning of its main sequence evolution (39.25$\,\mathrm{M}_{\odot}$). After
the secondary exhausted all hydrogen in the core, it contracts and
increase its luminosity and effective temperature. Due to the increase
of temperature in the envelope, the star ignites hydrogen in a shell,
expands and cools down drastically (dash-dotted line Fig.~\ref{secon},
$R$$\approx$2300$\rm~R_{\odot}$, $T_{\rm {eff}}$$\approx$3500$\rm~K$).
Meanwhile, the core contracts further and the core temperature
increases. Helium core burning starts when $T_{\rm
c}$$\approx$1.4$\cdot$10$^8 \rm~K$.  We stopped detailed modeling
when the secondary has $\sim$67\% of helium left in the core.

\begin{figure}
  \centering
   \includegraphics[width=\columnwidth]{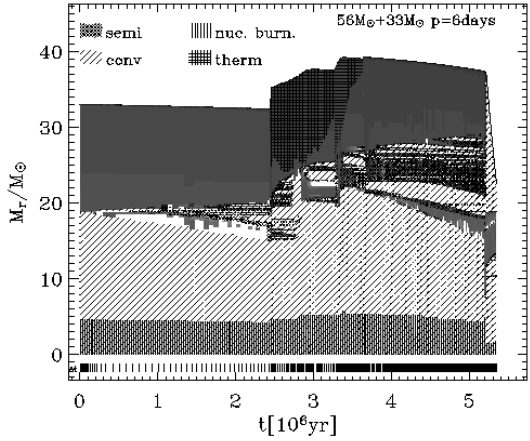}
  \caption[]{\label{evol}Evolution of the internal structure of a
  rotating 33$\,\mathrm{M}_{\odot}$ secondary from the ZAMS until red
  supergiant phase $Y_{\rm c}$=0.67.  Convection is indicated with
  diagonal hatching, semiconvection with crossed hatching and
  thermohaline mixing with straight crossed hatching.  The hatched
  area at the bottom indicates nuclear burning.  Gray shaded areas
  represent regions with rotationally induced mixing (intensity is
  indicated with different shades, the darker the colour, the stronger
  rotational mixing). The topmost solid line corresponds to the
  surface of the star.}
\end{figure}

\begin{figure}
  \centering
   \includegraphics[width=\columnwidth]{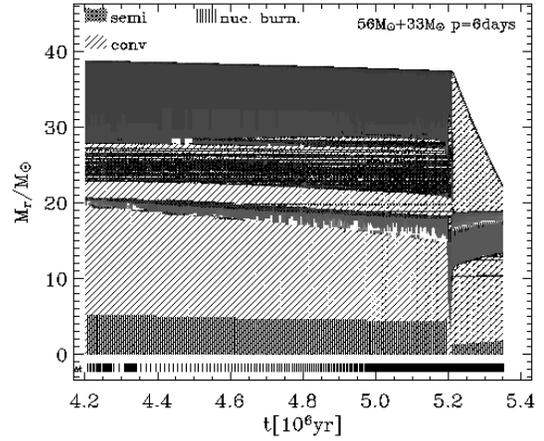}
  \caption[]{\label{sec}Evolution of the internal structure of a
  rotating 33$\,\mathrm{M}_{\odot}$ secondary from SN explosion of the
  primary and disruption of the system.  See Fig.~\ref{evol} for an explanation of the
  different hatching types.}
\end{figure}

Fig.~\ref{evol} shows the evolution of the internal structure of the
secondary from the ZAMS, through Case A and Case AB mass transfer and
the red supergiant phase until $Y_{\rm c}$=0.67.  The secondary starts
its main sequence evolution as a 33$\,\mathrm{M}_{\odot}$ star with a
convective core of $\sim$20$\,\mathrm{M}_{\odot}$.  Rotationally induced
mixing processes take place in the radiative envelope of the star.
The secondary then accretes $\sim$3$\,\mathrm{M}_{\odot}$ during the fast
phase and $\sim$4$\,\mathrm{M}_{\odot}$ during the slow phase of Case A mass
transfer.  Heavier elements accreted on the surface are relocated by
thermohaline mixing process and the convective core increases its mass
($\sim$25$\,\mathrm{M}_{\odot}$).  The secondary becomes a rejuvenated
$\sim$39$\,\mathrm{M}_{\odot}$ core hydrogen burning star in a WR+O binary
system. The system is likely to be disrupted by the explosion of the
primary while the secondary is still a main sequence star.  Details of
the evolution of the internal structure of the secondary after the
primary exploded are shown in Fig.~\ref{sec}.  After hydrogen is
exhausted in the core, the secondary becomes a $\sim$37$\,\mathrm{M}_{\odot}$
red supergiant with central helium burning in convective core of
$\sim$12$\,\mathrm{M}_{\odot}$ and a convective envelope of $\sim$20$\,\mathrm
{M}_{\odot}$. Due to stellar wind mass loss of $\sim$10$^{-4} \,\mathrm
M_{\odot} \rm~yr^{-1}$, the mass of the envelope rapidly decreases and
reaches $\sim$10$\,\mathrm{M}_{\odot}$ at the end of our calculations.

\begin{figure}
  \centering
   \includegraphics[width=\columnwidth]{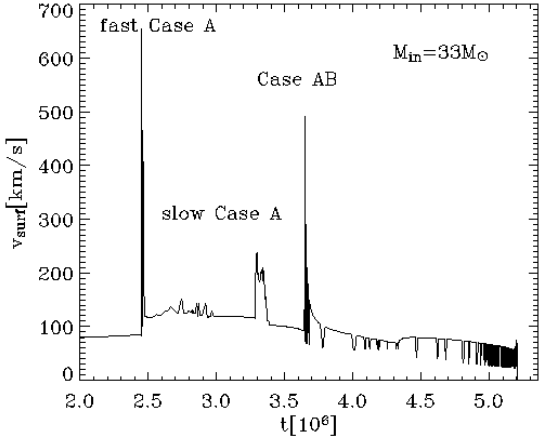}
  \caption[]{\label{vsurf}Surface rotational velocity of the secondary
  star.  Matter transfered from the primary with high mass transfer
  rate $\sim$10$^{-3}$-10$^{-4} \,\mathrm{M}_{\odot} \rm~yr^{-1}$ during the fast
  phase of Case A and Case AB spins up the surface layers of the
  secondary up to 500-700$\rm~km~s^{-1}$. During the slow phase of
  Case A mass transfer rate is lower ($\sim$10$^{-6} \,\mathrm
  {M}_{\odot} \rm~yr^{-1}$) and surface rotational velocity increases up to
  $\sim$200$\rm~km~s^{-1}$.  After Case AB mass transfer the secondary
  synchronizes with the orbital motion in the WR+O binary system.
  After the SN explosion of the primary, the secondary star evolves
  into a red supergiant with slowly rotating envelope of
  $\sim$0.02$\rm~km~s^{-1}$.}
\end{figure}

The modeled binary system starts its evolution with both components
synchronized to the orbital rotation. The surface rotational velocity
of the secondary star is $\sim$64$~\rm km~s^{-1}$. When the fast Case
A mass transfer starts, the secondary accretes matter from the primary
with high mass transfer rates ($\sim$10$^{-3} \,\mathrm
{M}_{\odot} \rm~yr^{-1}$). This matter carries angular momentum and spins up
the top layers of the secondary star (Fig.~\ref{vsurf}).  The mass
transfer rate during the slow phase of Case A is significantly lower
(10$^{-6}\,\mathrm{M}_{\odot} \rm~yr^{-1}$) and the surface rotational velocity of
the secondary increases to about 200$~\rm km~s^{-1}$.  For Case AB we
assumed a mass transfer rate of $\sim$10$^{-4} \,\mathrm{M}_{\odot} \rm~yr^{-1}$
(Sect.~\ref{comp}) and this spins up the surface of the secondary to
$\sim$500$\rm~km~s^{-1}$.  When the secondary spins up to close to
critical rotation it loses more mass according to Eq.~2 .
High mass loss decreases the net accretion efficiency and also removes
angular momentum from the secondary star. The secondary star is also
spun-down by tidal forces that tend to synchronize it with the orbital
motion \citep{2004AAinprep}.  After Case AB mass transfer, the
secondary will synchronize with the orbital motion in the WR+O binary
system.  After the SN explosion of the primary, the secondary becomes
a red supergiant with a very slowly rotating envelope and its surface
rotational velocity drops to less than 0.02$~\rm km~s^{-1}$.

\begin{figure}
  \centering
   \includegraphics[width=\columnwidth]{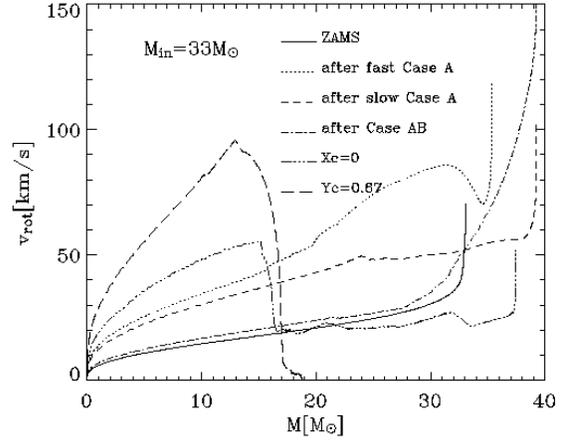}
  \caption[]{\label{vrot}Rotational velocity profiles of the secondary
  star on the hydrogen ZAMS (solid line), after the fast (dotted line)
  and the slow (short dashed line) phase of Case A mass transfer,
  after Case AB mass transfer (dash-dotted line), when helium ignites
  in the core (three dots-dashed line) and when the central helium
  abundance is 67\% (long dashed-line). }
\end{figure}

Fig.~\ref{vrot} shows rotational velocity profiles of the secondary
star in different evolutionary phases.  The surface of the secondary
gains additional angular momentum during mass transfer, as we already
explained.  This angular momentum is transported through the stellar
interior, and into the core.  The result is that the stellar core
spins faster on the helium ZAMS (Fig.~\ref{vrot}, three dots-dashed
line) than at the beginning of hydrogen core burning (Fig.~\ref{vrot},
solid line).  After this, the core contracts and further
increases its rotational velocity, and the envelope expands and
slows down.  When the central abundance of helium is around 67\%, the
maximum rotational velocity of the core is $\sim$100$~\rm km~s^{-1}$.

\begin{figure}
  \centering
   \includegraphics[width=\columnwidth]{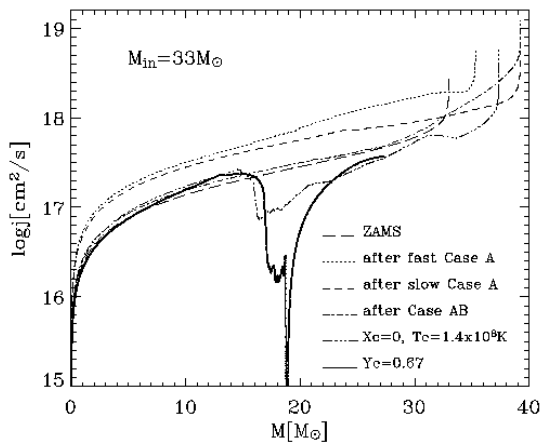}
  \caption[]{\label{core}Specific angular momentum profiles of the
  secondary star on the hydrogen ZAMS (long dashed line), after fast
  (dotted line) and slow (short dashed line) Case A mass transfer,
  after Case AB mass transfer (dash-dotted line), when helium ignites
  in the core (three dots-dashed line) and when the central helium
  abundance is 67\% (solid line).}
\end{figure}

Fig.~\ref{core} shows specific angular momentum profiles of the
secondary at different points of its evolution. The specific angular
momentum of the secondary increases significantly due to fast Case A
mass transfer (Fig.~\ref{core}, dotted line).  After this, the
secondary loses angular momentum due to stellar wind mass loss, but
also gains certain amount through slow Case A and Case AB mass
transfer (Fig.~\ref{core}, dashed and dot-dashed line).  The result is
that the core has a larger specific angular momentum when central
helium burning starts than at the beginning of hydrogen core burning.
After core hydrogen exhaustion, the secondary evolves into a red
supergiant, the core contracts and the envelope expands.  This leads
to a spin-up of the core and a spin-down of the envelope.  The
specific angular momentum of the core at 3$\,\mathrm{M}_{\odot}$ is
$\sim$5.5$\cdot$10$^{16} \rm~cm^2~s^{-1}$ (Fig.~\ref{core}, three
dot-dashed line).  The envelope is convective and slowly rotating
($\sim$0.02$~\rm km~s^{-1}$). The core is rigidly rotating with
maximum rotational velocity of $\sim$100$~\rm km~s^{-1}$.  The core
and the envelope are separated by layers that have a high
$\mu$-gradient.  Angular momentum is not efficiently transported
through these layers, so the core is not slowed down by the slow
rotation of the envelope.  When a third of the central helium supply
is exhausted, the core (at 3$\,\mathrm{M}_{\odot}$) has a specific angular
momentum of $\sim$5$\cdot$10$^{16} \rm~cm^2~s^{-1}$.  If we assume
that the angular momentum of the core decreases further during helium
core burning with the same rate, specific angular momentum of the core
at the moment of helium exhaustion is expected to be
$\sim$4$\cdot$10$^{16}\rm~cm^2~s^{-1}$.  As we have already seen for
single stars, there is no angular momentum loss from the core during
core carbon burning, so we can conclude that this star has enough
angular momentum to produce a collapsar and, in the case that the
hydrogen envelope is lost during red supergiant phase, a gamma-ray
burst.

\begin{figure}
  \centering
   \includegraphics[width=\columnwidth]{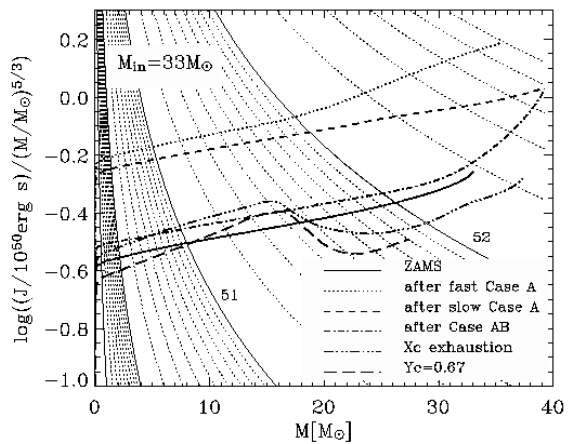}
  \caption[]{\label{jt}Logarithm of the integrated angular momentum
  $J(m)$=$\int_{0}^{M}j(m^{\prime})dm^{\prime}$ divided by $m^{5/3}$,
  as a function of the mass coordinate $m$ for a 33$\,\mathrm{M}_{\odot}$
  secondary star for the same evolutionary stages as shown in
  Fig.~\ref{core}. The thin lines give a logarithmic scale of levels
  of constant $J$ labeled with $\rm log(J/\rm (erg~s))$.}
\end{figure}

\begin{figure}
  \centering
   \includegraphics[width=\columnwidth]{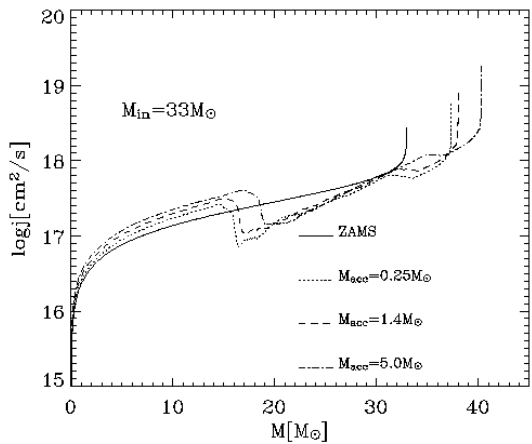}
  \caption[]{\label{core2}Specific angular momentum profiles of the
  secondary star on the hydrogen ZAMS (solid line) and when helium
  ignites in the core ($T_c$=1.4$\cdot$10$^8\rm~K$), assuming the
  accretion of 0.25$\,\mathrm{M}_{\odot}$ (dotted line) 1.4$\,\mathrm{M}_{\odot}$
  (dashed line) or 5$\,\mathrm{M}_{\odot}$ (dot-dashed line) during Case AB
  mass transfer.}
\end{figure}

The gain of angular momentum of an accreting star is proportional to
the amount of accreted matter.  We show in Fig.~\ref{core2} specific
angular momentum profiles of the secondary ($X_c$=0,
$T_c$=1.4$\cdot$10$^8 \rm~K$) assuming the accretion of 0.25, 1.4 and
5$\,\mathrm{M}_{\odot}$ during Case AB mass transfer.  We see that the
specific angular momentum of the core is higher if the star accreted
more matter during mass transfer. The specific angular momentum of the
core at 3$\,\mathrm{M}_{\odot}$ increases from 6$\cdot$10$^{16} \rm~cm^2~s^{-1}$ to 
8$\cdot$10$^{16} \rm~cm^2~s^{-1}$.

\section{Models with magnetic field}\label{magnetic}
\subsection{Single stars}\label{magsingle}
\begin{figure}
  \centering
   \includegraphics[width=\columnwidth]{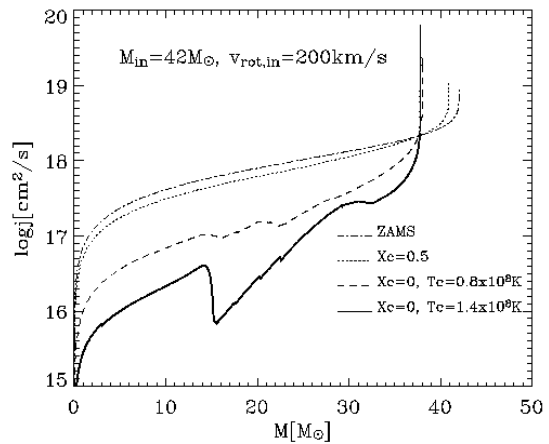}
  \caption[]{\label{coremag}Specific angular momentum profiles of the
  42$\,\mathrm{M}_{\odot}$ single star with magnetic fields on the hydrogen
  ZAMS (dot-dashed line), when 50\% of hydrogen is left in the
  center (dotted line), when hydrogen is exhausted in the core (dashed
  line) and when central helium burning starts (solid line).}
\end{figure}

We model the evolution of a single 42$\,\mathrm{M}_{\odot}$ star with an
initial rotational surface velocity of 200$~\rm km~s^{-1}$ and
magnetic field included.  Fig.~\ref{coremag} shows specific angular
momentum profiles in different phases of the evolution. We see that
the initial (ZAMS) profiles of the star without magnetic field
(Fig.~\ref{core42}, dot-dashed line) and the star with magnetic field
(Fig.~\ref{coremag}, dot-dashed line) are identical and represent
solid body rotation.  We notice, however, that the magnetic star loses
more angular momentum from the core during the main sequence evolution
as well as between hydrogen core exhaustion (Fig.~\ref{coremag},
dashed line) and core helium ignition (Fig.~\ref{coremag}, solid
line).

\begin{figure}
  \centering
   \includegraphics[width=\columnwidth]{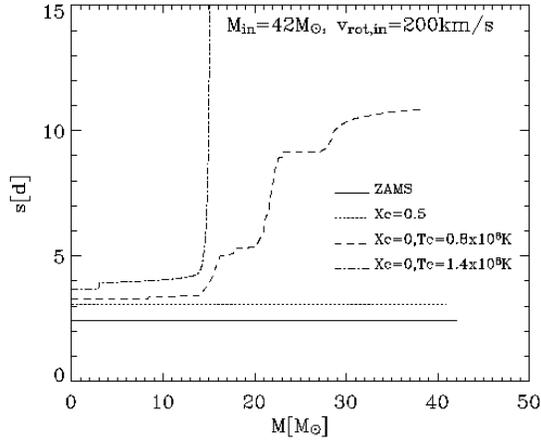}
  \caption[]{\label{magomega}Spin period profiles of the 42$\,\mathrm
  {M}_{\odot}$ single star on the ZAMS (solid line), when 50\%
  hydrogen is left in the center (dotted line), when hydrogen is
  exhausted in the core (dashed line), and when central helium burning
  starts (dot-dashed line).}
\end{figure}

\begin{figure}
  \centering
   \includegraphics[width=\columnwidth]{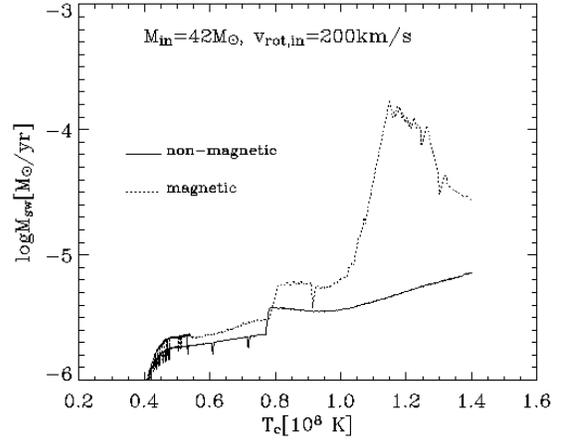}
  \caption[]{\label{magml}Stellar wind mass loss of the magnetic
  (dotted line) and non-magnetic (solid line) 42$\,\mathrm{M}_{\odot}$ star.
  The star with magnetic fields loses more mass because magnetic
  torques transport angular momentum, which spins-up the surface
  layers and enhances mass loss. }
\end{figure}

The star loses mass by a stellar wind. This removes angular momentum
from the surface layers and slows them down.  Since the magnetic
torque keeps the star close to solid body rotation during the main
sequence evolution (Fig.~\ref{magomega}), however, angular momentum is
transported from the stellar interior towards the surface. The
viscosity due to the magnetic field is a few orders of magnitude
larger than the one for rotational mixing ($\sim$10$^{{10}-{12}}\rm~
cm^2~s^{-1}$ compared with $\sim$10$^{{7}-{8}}\rm~cm^2~s^{-1}$), and
it can overcome the $\mu$-gradient barrier.  Therefore, the surface of
the star is spun up and the core is slowed down.  The surface layers
rotate faster than in the corresponding non-magnetic star, and the
stellar wind mass loss is higher (Fig.~\ref{magml}).

Summarizing, the core of a magnetic star loses significantly more
angular momentum during the main sequence evolution than when magnetic
fields are not included.  One reason is that the $\mu$-gradient does
not represent a strong barrier for angular momentum transport. And
second, because the surface layers are spun-up, the stellar wind mass
loss is larger and this requires more angular momentum from the core
to keep solid body rotation of the star.  When hydrogen burning is
finished ($T_c$=0.8$\cdot 10^8 \rm~K$), the specific angular momentum
of the core (at 3$\,\mathrm{M}_{\odot}$) is less than 3$\cdot10^{16} \rm~cm^2~s^{-1}$ 
(Fig.~\ref{coremag}, dashed line).

When the star expands after core hydrogen burning, magnetic torques
still work toward enforcing solid body rotation.  The envelope
expansion, however, is too fast and angular momentum transport is not
efficient enough, to keep the envelope synchronized with the core
(Fig.~\ref{magomega}).  Since magnetic viscosity can overcome the
$\mu$-gradient, the core is slowed down and the envelope is spun up.
Stellar wind mass loss is enhanced due to faster rotation and is
significantly higher than in the case of the non-magnetic star
(Fig.~\ref{magml}).  The star loses significant an amount of angular
momentum between hydrogen exhaustion and helium ignition and the
specific angular momentum of the core (at 3$\,\mathrm{M}_{\odot}$) at the
time helium core burning starts is only 5$\cdot10^{15} \rm~cm^2~s^{-1}$.

\begin{figure}
  \centering
   \includegraphics[width=\columnwidth]{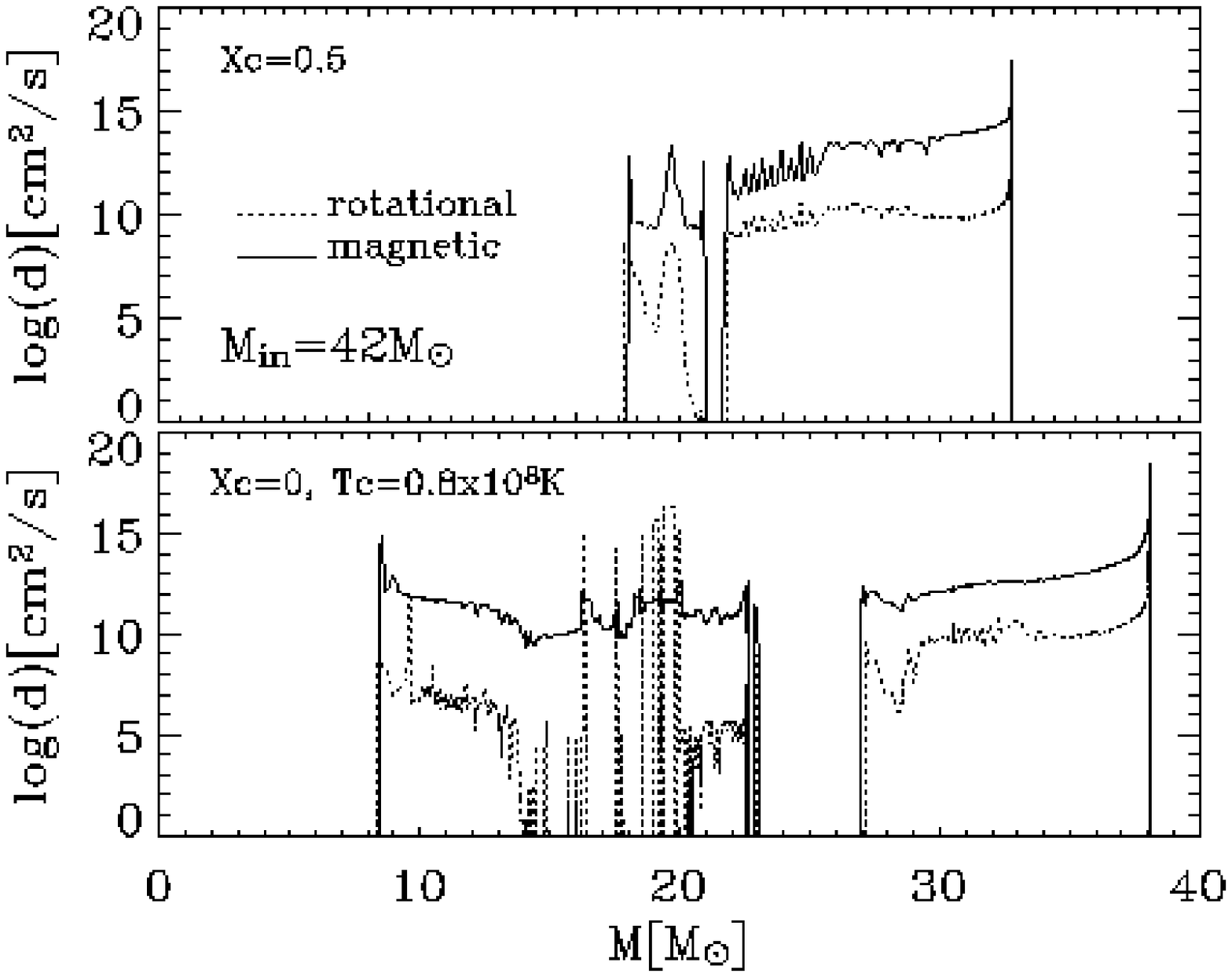}
  \caption[]{\label{diff}Effective viscosities 
  due to magnetic torques (solid line) and due to rotational
  instabilities (dotted line) for a rotating 42$\,\mathrm{M}_{\odot}$ star at
  a central hydrogen abundance of 50\% and at core hydrogen
  exhaustion.}
\end{figure}

\subsection{Binary systems}\label{magbinary}
We modeled the evolution of a rotating binary system with magnetic
fields, initial masses $M_{\rm 1,in}$=56$\,\mathrm{M}_{\odot}$ and $M_{\rm
2,in}$=33$\,\mathrm{M}_{\odot}$, and an initial orbital period of $p_{\rm
in}$=6 days.  The binary system starts synchronized on the ZAMS with
initial surface rotational velocities of $\sim$90$~\rm km~s^{-1}$ for
the primary and $\sim$60$~\rm km~s^{-1}$ for the secondary which is,
as we already mentioned for non-magnetic models, much slower than the
typical values for single stars of these masses.

During the fast phase of Case A mass transfer ($\dot
M_{tr}$$\sim$10$^{-3} \,\mathrm{M}_{\odot} \rm~yr^{-1}$), the primary loses
$\sim$20$\,\mathrm{M}_{\odot}$ and the secondary accretes only about
4.5$\,\mathrm{M}_{\odot}$ of that matter due to the mass loss caused by rotation
close to break-up.  After the fast process of Case A mass transfer,
the primary continues to expand on a nuclear time scale and to
transfer mass onto the secondary star, but with much lower mass
transfer rates (slow Case A, $\dot M_{tr}$$\sim$10$^{-6} \,\mathrm
{M}_{\odot} \rm~yr^{-1}$).  At the end of core hydrogen burning the primary
contracts and thus RLOF stops.

Accretion increases rotational velocity of the surface of the
secondary star to almost 400$~\rm km~s^{-1}$ during the fast phase of
Case A mass transfer.  The mass transfer rate during slow Case A mass
transfer is significantly lower (10$^{-6}\,\mathrm{M}_{\odot} \rm~yr^{-1}$), and
the surface rotational velocity of the secondary increases to about
200$~\rm km~s^{-1}$.

\begin{figure}
  \centering
   \includegraphics[width=\columnwidth]{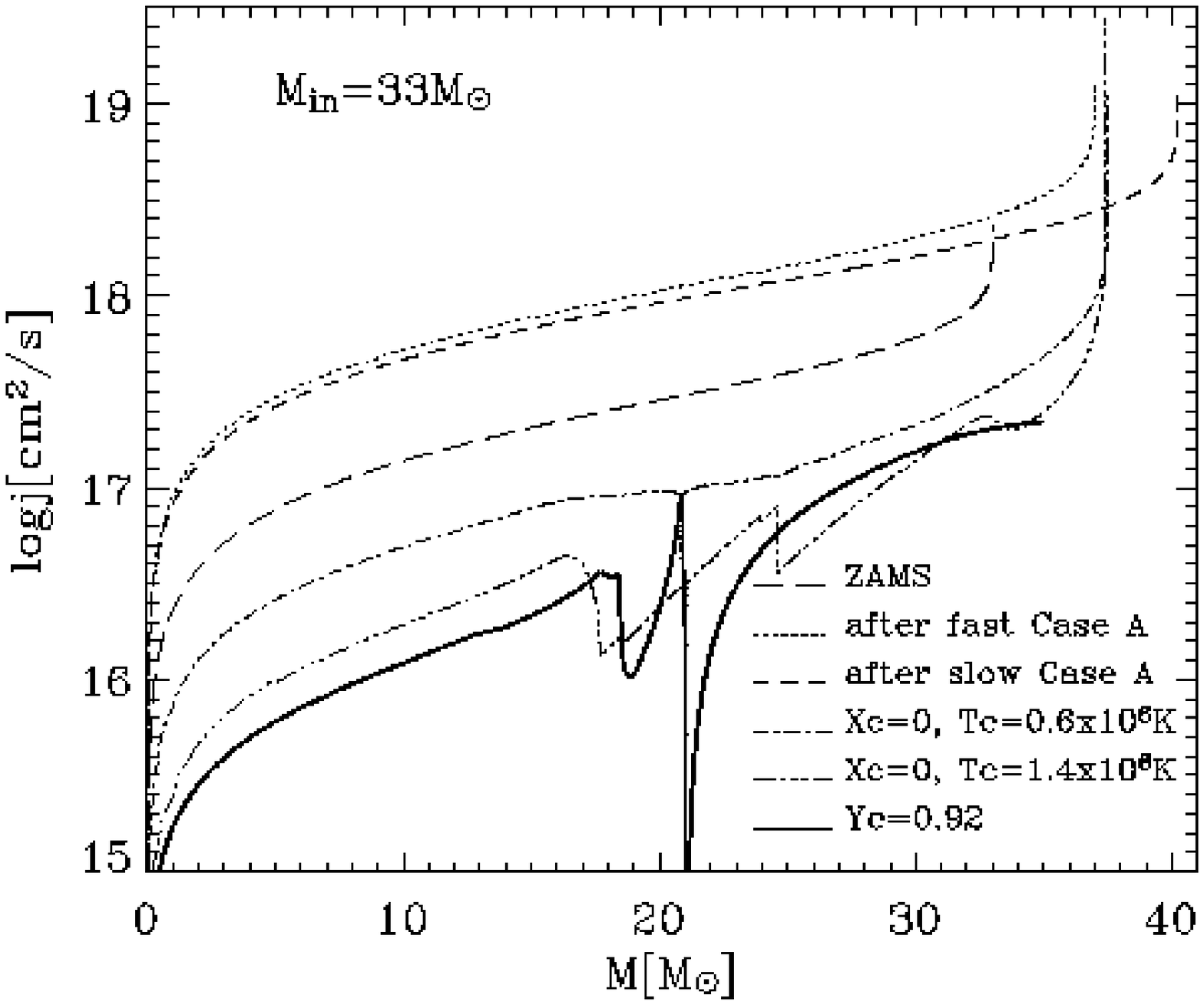}
  \caption[]{\label{magbincore}Specific angular momentum profiles of
  the secondary star with magnetic fields on the hydrogen ZAMS (long
  dashed line), after fast (dotted line) and slow (short dashed line)
  Case~A mass transfer, when all hydrogen is exhausted in the center
  (dot-dashed line), when helium ignites (three dots-dashed line) and
  and when the central helium abundance is 92\% (solid line).}
\end{figure}

Fig.~\ref{magbincore} shows specific angular momentum profiles of the
secondary at different phases of evolution.  The specific angular
momentum of the secondary increases significantly due to the fast Case
A mass transfer (Fig.~\ref{magbincore}, dotted line).  Angular
momentum is transported more efficiently through the stellar interior
compared to the non-magnetic model, since the
incurred magnetic torques are a few orders of magnitude more efficient
in angular momentum transport than the rotational instabilities 
(Fig.~\ref{diff}).  Comparing the specific angular momentum of
the non-magnetic (Fig.~\ref{core}) and magnetic model
(Fig.~\ref{magbincore}), we notice that during fast Case A the angular
momentum of the magnetic star increases more than that of the
corresponding non-magnetic star (2$\cdot$10$^{17}\rm~cm^2~s^{-1}$ for
magnetic and 1.25$\cdot$10$^{17}\rm~cm^2~s^{-1}$ for non-magnetic
star, $\sim$10$^4\rm~yrs$ after fast Case A, at 3$\,\mathrm{M}_{\odot}$).

The accretion stops when the secondary still has almost 50\% of the
hydrogen to burn in the core.  Angular momentum is efficiently
transported from the stellar core to the surface and the
$\mu$-gradient can not stop it as in the case of the non-magnetic
star.  During further main sequence evolution, the stellar core loses
significant angular momentum and when hydrogen core burning stops, the
specific angular momentum at 3$\,\mathrm{M}_{\odot}$ is 2$\cdot$10$^{16}\rm
~cm^2~s^{-1}$. Before helium ignites in the core, the specific angular
momentum decreases to 6$\cdot$10$^{15}\rm~cm^2~s^{-1}$.  The further
evolution through the red supergiant phase slows down the core, since
the envelope slows it down (magnetic viscosity can overcome the
$\mu$-barrier).

\begin{figure}
  \centering
   \includegraphics[width=\columnwidth]{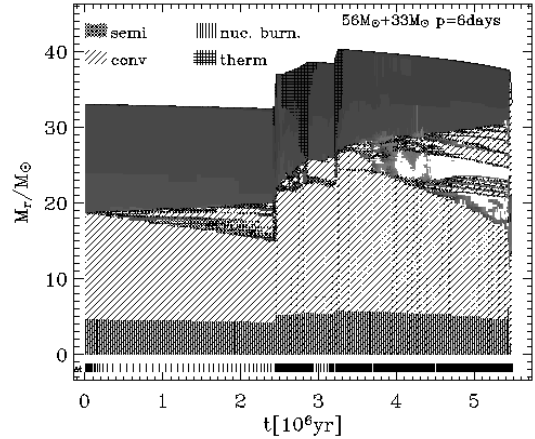}
  \caption[]{\label{magconv}Evolution of the internal structure of a
  rotating 33$\,\mathrm{M}_{\odot}$ secondary star with magnetic field from
  the ZAMS until red supergiant phase $Y_{\rm c}$=0.92.
  See Fig.~\ref{evol} for an explanation of
  the different hatching types.}
\end{figure}

\begin{figure}
  \centering
   \includegraphics[width=\columnwidth]{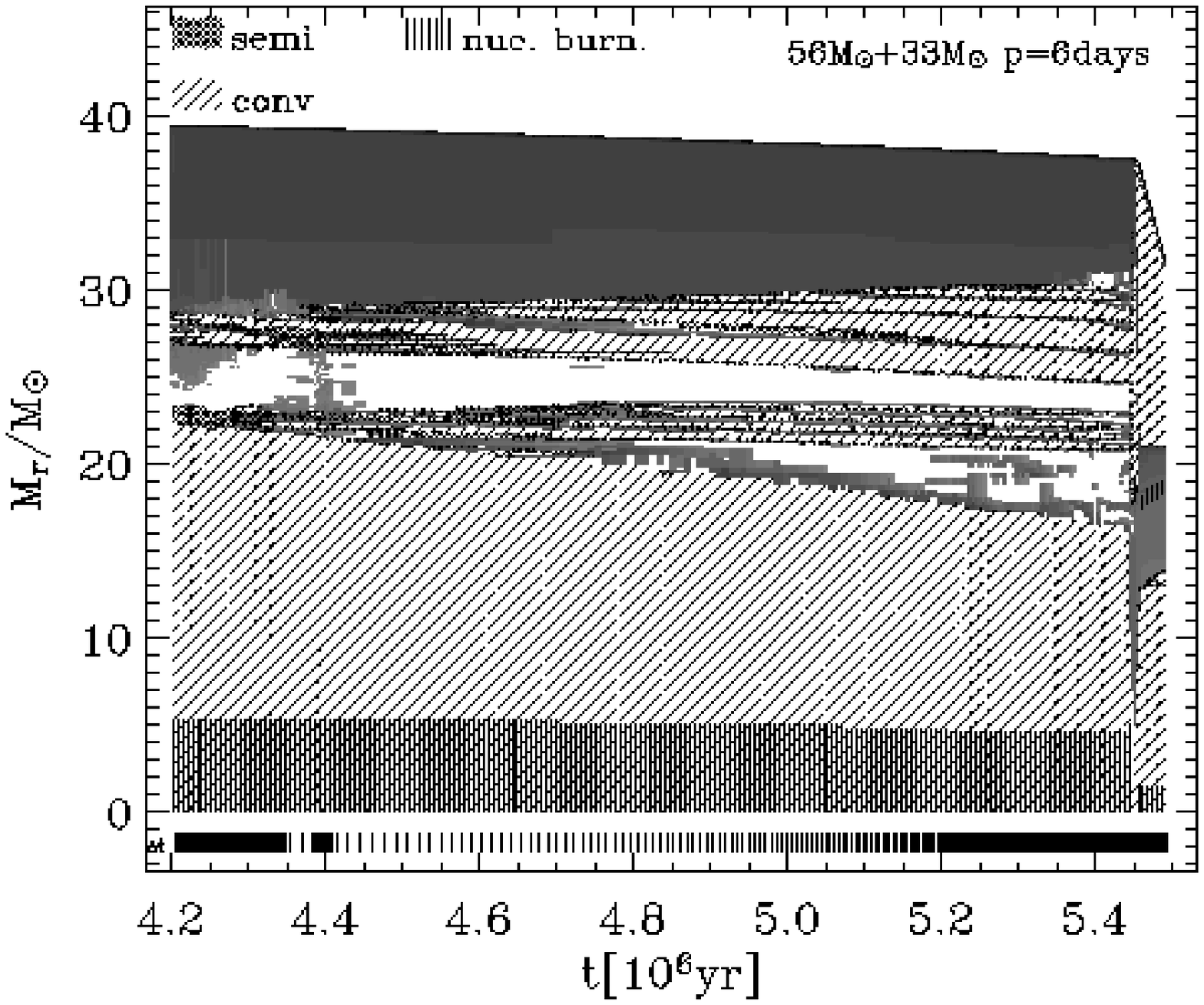}
  \caption[]{\label{magconv}Evolution of the internal structure of a
  rotating 33$\,\mathrm{M}_{\odot}$ secondary star with magnetic field from
  after the SN explosion of the primary and disruption of the system.
  See Fig.~\ref{evol} for an explanation of
  the different hatching types.}
\end{figure}

Apparently, if a star with magnetic fields accretes matter half way in
the main sequence evolution, it has enough time to lose angular
momentum before hydrogen is exhausted in the core.  This is why we
compute models of non-magnetic and magnetic 33$\,\mathrm{M}_{\odot}$ stars
that accrete 2$\,\mathrm{M}_{\odot}$ at the end of their main sequence
evolution ($Y_c$=0.95).  We assume an initial surface velocity of
these stars of $\sim$60$~\rm km~s^{-1}$ since they evolve in a binary
system.

\begin{figure}
  \centering
   \includegraphics[width=\columnwidth]{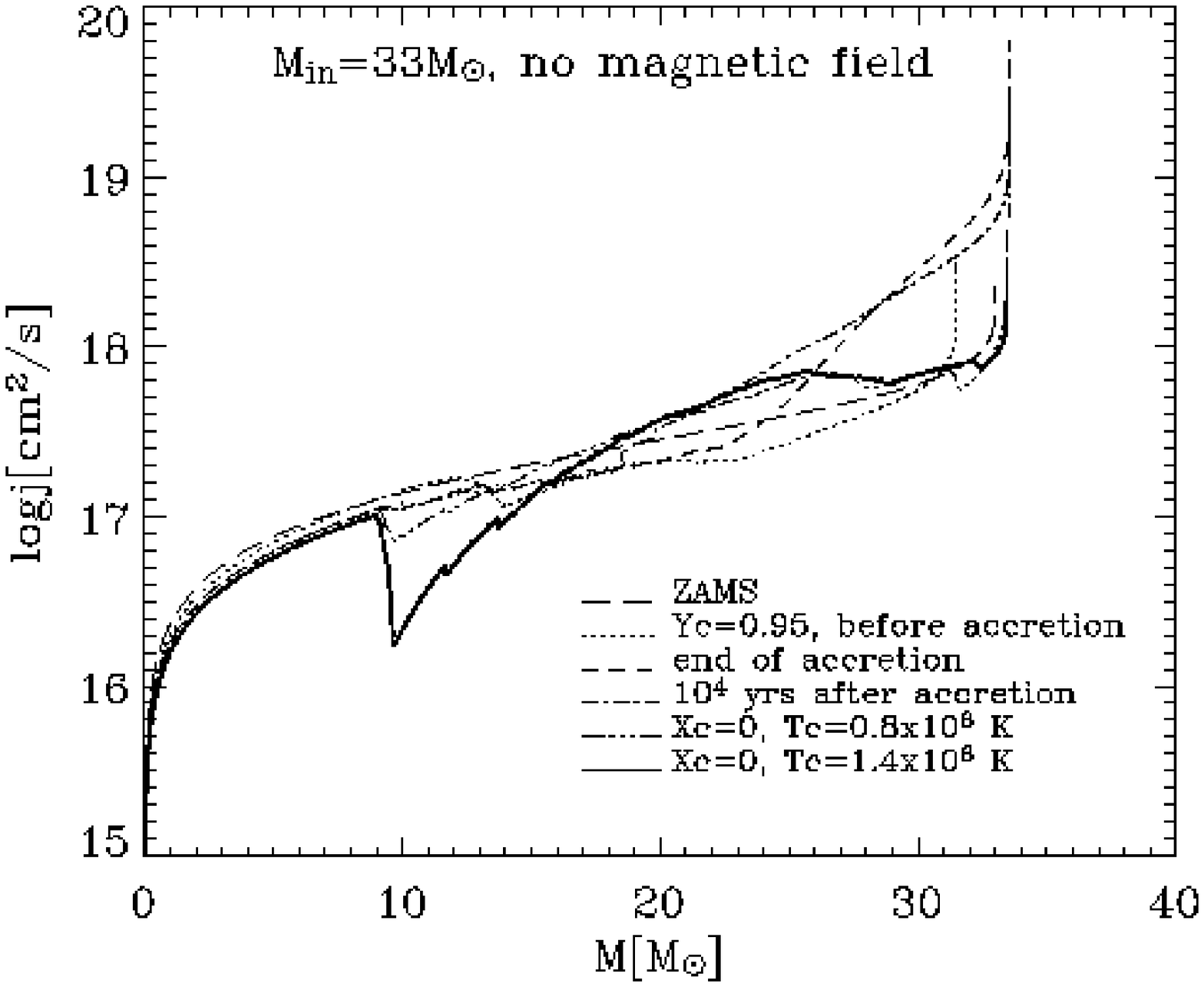}
  \caption[]{\label{acccore}Specific angular momentum profiles of a
  33$\,\mathrm{M}_{\odot}$ non-magnetic star on the ZAMS (long dashed line),
  when $Y_c$=95\% (dotted line), at the end of the accretion (dashed
  line), 10$^4\rm~yrs$ after the accretion (dot-dashed line), when
  hydrogen is exhausted in the core (three dot-dashed line) and when
  helium ignites in the core (solid line).}
\end{figure}

\begin{figure}
  \centering
   \includegraphics[width=\columnwidth]{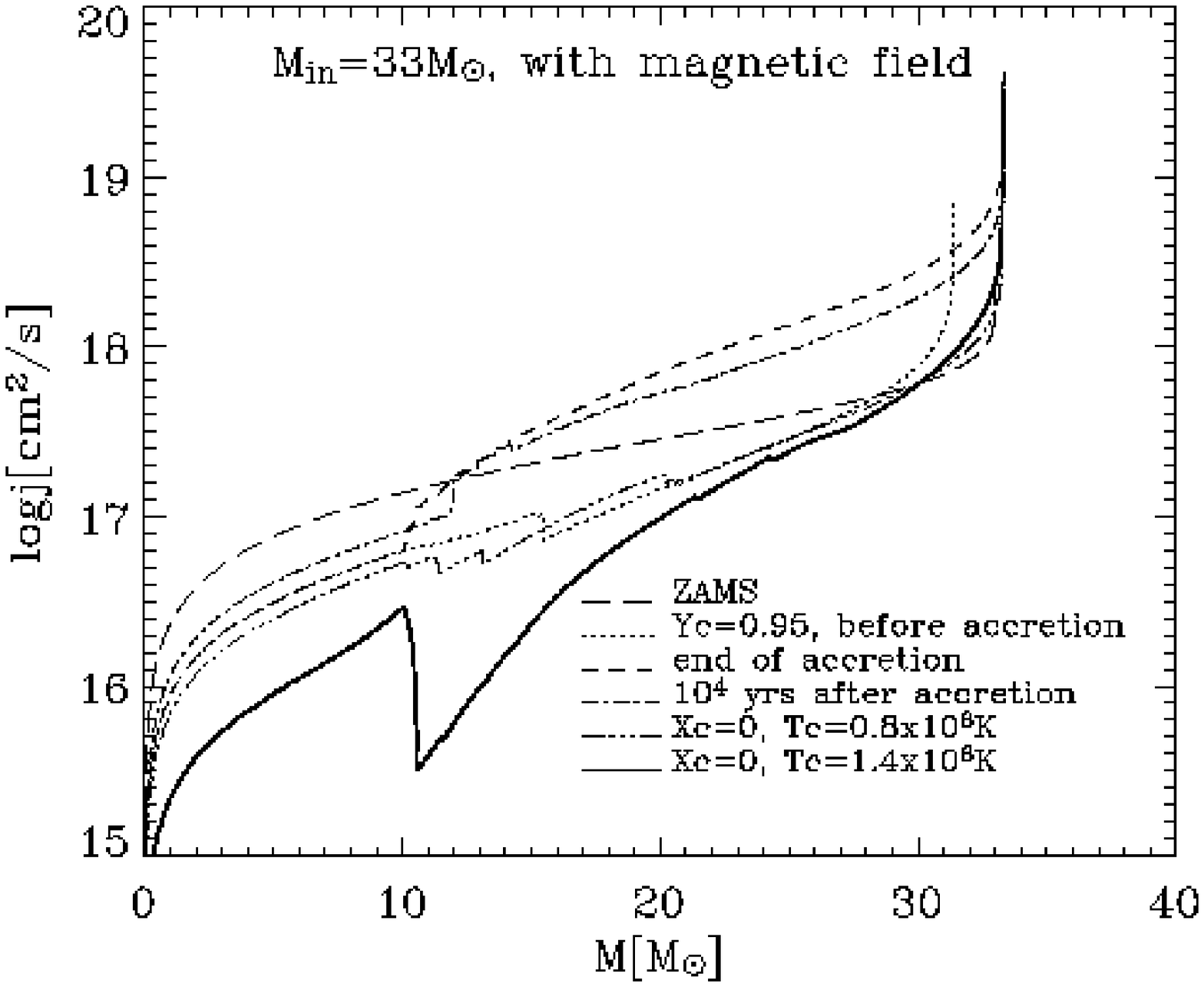}
  \caption[]{\label{accmagcore}Specific angular momentum profiles of a
  33$\,\mathrm{M}_{\odot}$ magnetic star on the ZAMS (long dashed line), when
  $Y_c$=95\% (dotted line), at the end of the accretion (dashed line),
  10$^4\rm~yrs$ after the accretion (dot-dashed line), when hydrogen
  is exhausted in the core (three dot-dashed line) and when helium
  ignites in the core (solid line).}
\end{figure}

When the central helium abundance reaches 95\%, the core of the 33$\,\mathrm
{M}_{\odot}$ star with magnetic fields has a specific angular momentum
of about 2$\cdot$10$^{16}\rm~cm^2~s^{-1}$ (Fig.~\ref{accmagcore},
dotted line).  On the other hand, the core of a star without magnetic
fields hardly loses any angular momentum and its specific angular
momentum is 4$\cdot$10$^{16}\rm~cm^2~s^{-1}$ (Fig.~\ref{acccore},
dotted line).  After this, both stars accrete 2$\,\mathrm{M}_{\odot}$ of
matter with an accretion rate of 10$^{-4}\,\mathrm{M}_{\odot} \rm~yr^{-1}$. We
assume that in this case stellar wind mass loss is not enhanced by
rotation, so there is no additional angular
momentum loss due to this effect. Accreted matter adds angular
momentum to the surface layers and this angular momentum is then
transported through the stellar interior much more efficiently in the
model with magnetic field.  At the end of accretion
(Fig.~\ref{acccore} and Fig.~\ref{accmagcore}, dashed line), angular
momentum has been transported further inward in the magnetic star than
in the non-magnetic star.  If we look at the specific angular momentum
profiles 10$^4\rm~yrs$ after the accretion ended, we can notice that
the specific angular momentum of the core of the magnetic star
increases to almost 3$\cdot$10$^{16}\rm~cm^2~s^{-1}$ (at 3$\,\mathrm
{M}_{\odot}$). On the other hand, in the
non-magnetic star angular momentum has been transported inward but
still has not reached the core. 

The magnetic star continues to lose angular momentum from the core
during the remaining main sequence evolution
($\sim$10$^5 \rm~yrs$), and when hydrogen is exhausted in the core the
specific angular momentum at 3$\,\mathrm{M}_{\odot}$ is 1.6$\cdot$10$^{16}\rm~
cm^2~s^{-1}$.  As explained above, in the
magnetic model, the core loses significant amounts of angular
momentum between core hydrogen exhaustion and core helium ignition,
and when central helium burning starts, the specific angular momentum
of the core at 3$\,\mathrm{M}_{\odot}$ is only 6$\cdot$10$^{15}\rm~cm^2~s^{-1}$.

\begin{figure}
  \centering
   \includegraphics[width=\columnwidth]{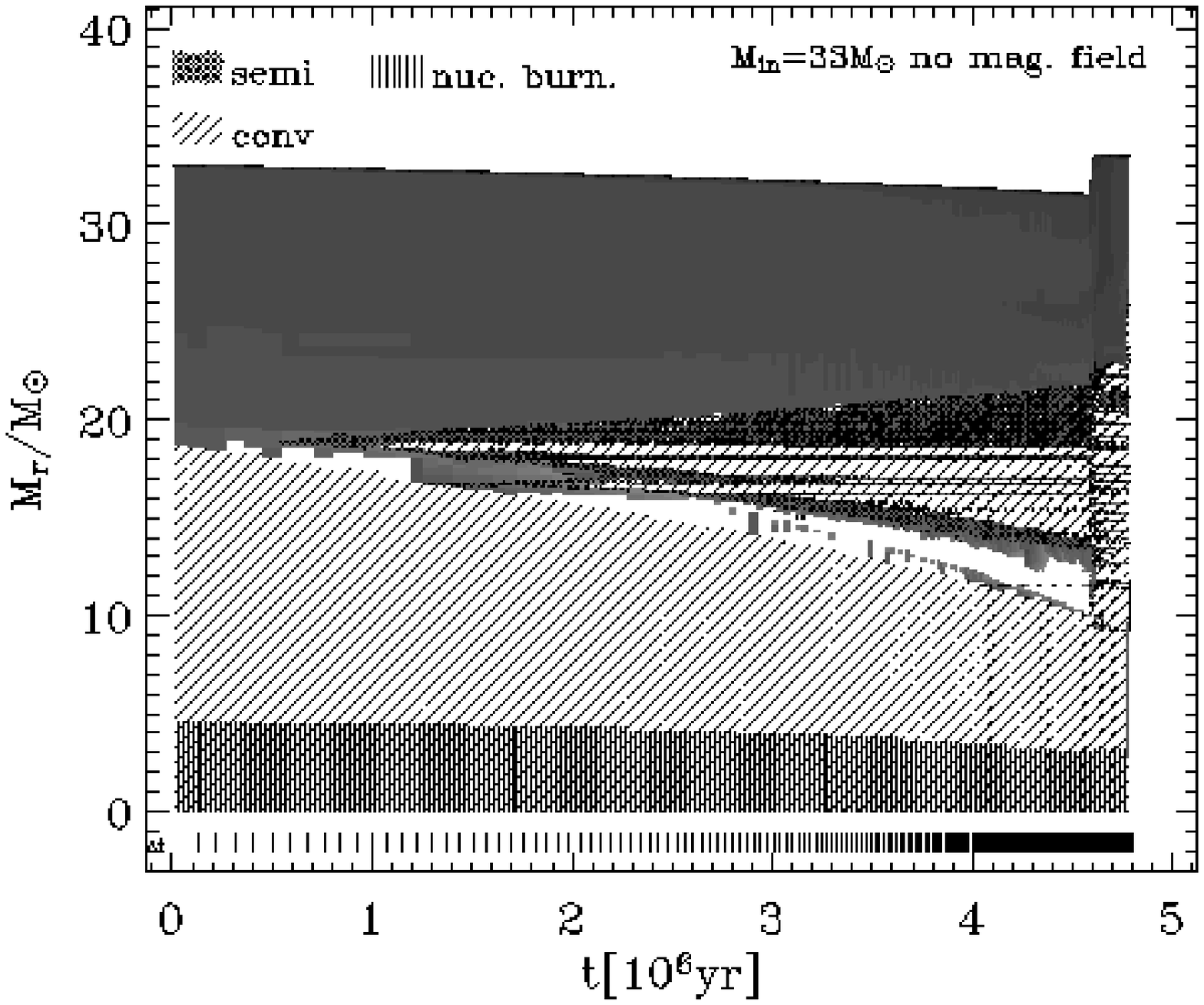}
  \caption[]{\label{accrotconv}Evolution of the internal structure of
  a rotating 33$\,\mathrm{M}_{\odot}$ secondary star until helium ignition in
  the core. The star accretes 2$\,\mathrm{M}_{\odot}$ when the central helium
  abundance is 95\%.  See Fig.~\ref{evol} for
  an explanation of the different hatching types.}
\end{figure}

\begin{figure}
  \centering
   \includegraphics[width=\columnwidth]{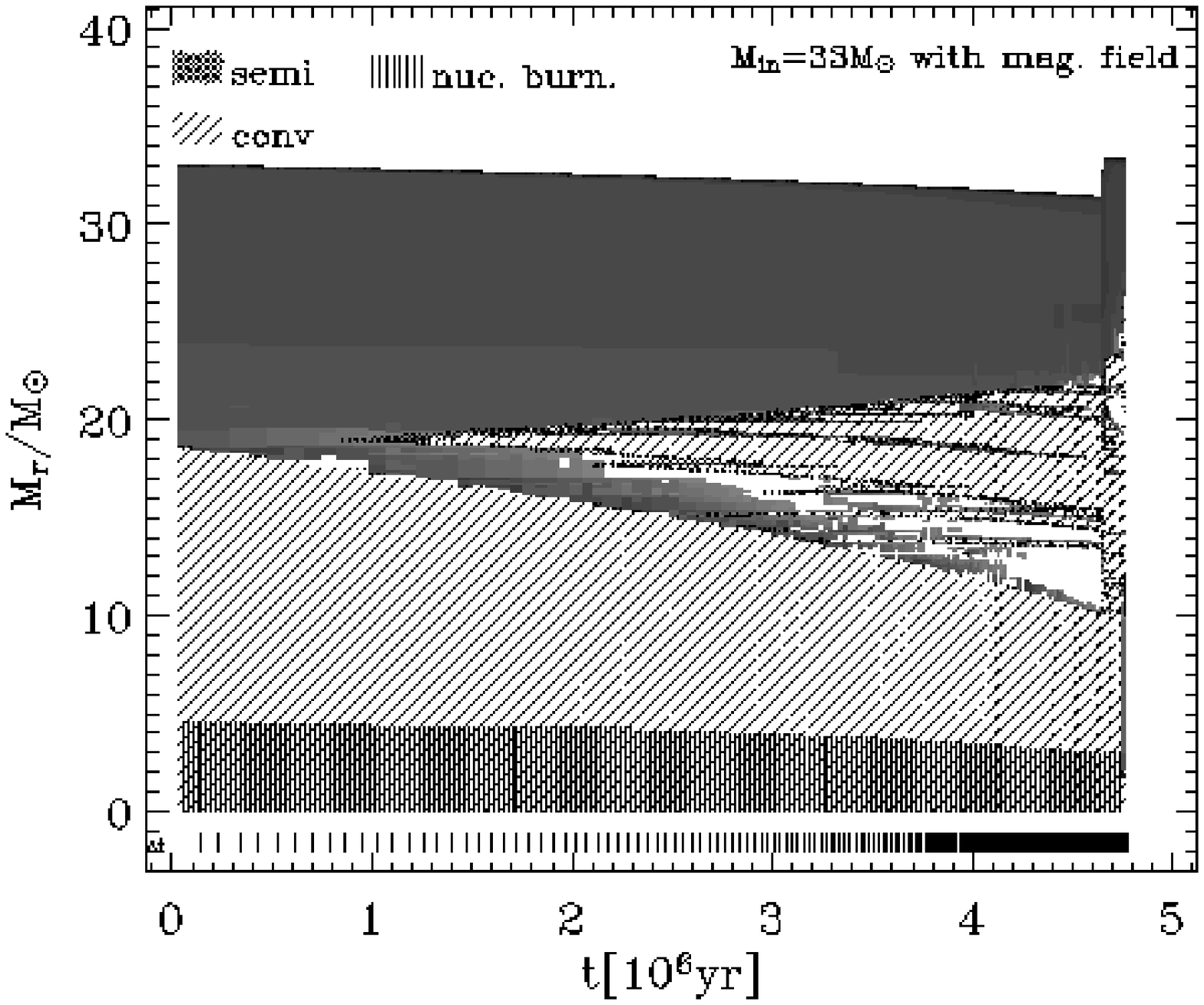}
  \caption[]{\label{accmagconv}Evolution of the internal structure of
  a rotating 33$\,\mathrm{M}_{\odot}$ secondary star with magnetic field
  until helium ignition in the core. The star accretes 2$\,\mathrm
  {M}_{\odot}$ when the central helium abundance is 95\%.
  See Fig.~\ref{evol} for an explanation of
  the different hatching types.}
\end{figure}

\section{Conclusions}\label{concl}

The single star models presented above confirm that non-magnetic stars
in the mass range relevant for GRB production through collapsars may
retain enough core angular momentum for a GRB to form. In particular,
the stars that evolve into red supergiants and become 
Wolf-Rayet stars only at the end of core helium burning can avoid
a significant core angular momentum loss through Wolf-Rayet winds and
may retain a specific angular momentum of up to 10$^{17}\rm~cm^2~s^{-1}$ 
in their iron cores.  In these models, an effective core
angular momentum loss during the main sequence is prevented by the
suppression of rotational mixing in regions containing a mean
molecular weight gradient.  Angular momentum loss from the stellar
core during helium burning is insignificant.  Contrary to the trend
found in \citet{2000ApJ...528..368H}, we find that a 42$\,\mathrm{M}_{\odot}$
star may end its life with a larger specific core angular momentum
than a 20$\,\mathrm{M}_{\odot}$.  The reason is that, for the same initial
equatorial rotation velocity, the initial specific angular momentum in
the 42$\,\mathrm{M}_{\odot}$ is significantly larger due to its larger
initial radius.  We note that a similar trend has been found by
\citet{2004A&A...425..649H}.

The dynamo model of \citet{2002A&A...381..923S} cause a significant
angular momentum transport even in the presence of mean molecular
weight gradients.  \citet{2004IAUS..215b} already found the increased
coupling of core and envelope to result in iron core specific angular
momenta of the order of 10$^{14}\rm~cm^2~s^{-1}$ for stars between 10
and 25$\,\mathrm{M}_{\odot}$.  Our magnetic 42$\,\mathrm{M}_{\odot}$ single star
model was only computed up to core helium ignition. By then, however,
the specific core angular momentum was already reduced by a factor of
30 from the initial value. \citet{2004IAUS..215b} showed that during
core helium burning the core-envelope coupling reduces the core
angular momentum by another order of magnitude.  The final specific
core angular momentum in our 42$\,\mathrm{M}_{\odot}$ star can thus be
estimated to fall below $\sim$10$^{15}\rm~cm^2~s^{-1}$, which will
render effects of rotation during the core collapse insignificant.

We also model the evolution of a 56$\,\mathrm{M}_{\odot}$+33$\,\mathrm{M}_{\odot}$
binary system with an initial orbital period of 6 days. Since the
binary-enhanced mass loss of the primary leads to extremely slow
rotation, our attention focuses on the secondary star, which is
spun-up due to accretion of mass and angular momentum.  We computed
two binary evolution sequences, with and without magnetic fields.  In
both sequences, the initial 33$\,\mathrm{M}_{\odot}$ star grows to about
40$\,\mathrm{M}_{\odot}$ due to accretion during its main sequence evolution,
and subsequently evolves into a red supergiant.  Before the mass
transfer, tidal spin-orbit coupling leads to bound rotation of both
stars and to a specific angular momentum which is a factor of
$3..5$ smaller than in a corresponding single star.

The accretion leads to a significant spin-up of the star. In the
non-magnetic model, the mean molecular weight gradient limits the
inward diffusion of angular momentum, and the corresponding spin-up of
the core merely compensates the tidally induced angular momentum
loss.  The result is a helium core that rotationally decouples from
the envelope with roughly the same mass and angular momentum as in the
case of a 42$\,\mathrm{M}_{\odot}$ single star, i.e., with high enough
specific angular momentum to produce a GRB.

In the magnetic model, the core spin-up due to accretion is stronger.
It temporarily leads to a core spin rate which is factor of $2..3$
above that of a ZAMS star of comparable mass.  Magnetic core-envelope
coupling, however, has reduced the specific core angular momentum by
almost a factor 100 by the time the star has started core helium
burning.  Its final core angular momentum will thus be comparable to
that of the magnetic 42$\,\mathrm{M}_{\odot}$ single star model discussed
above: probably too small to produce a GRB.

Clearly, the effect of core spin-up due to accretion on the final core
angular momentum will be larger if the accretion occurs later in the
evolution of the accretion star, as there will be less time to lose
again the angular momentum gained by accretion.  Therefore, we
performed the numerical experiment of accreting 2$\,\mathrm{M}_{\odot}$ of
material on a 33$\,\mathrm{M}_{\odot}$ at a time when the core hydrogen
concentration was down to 5\% (instead of about 40\% in the binary
evolution model). The result was again that the core-envelope coupling
in the magnetic model was overwhelming and prevents this scenario to
be a realistic option for GRB production.

We conclude that our binary models without magnetic field can
reproduce stellar cores with a high enough specific angular momentum
($j$$\ge$3$\cdot$10$^{16}\rm~cm^2~s^{-1}$) to produce a collapsar and
a GRB.

If magnetic field is taken into consideration, however, GRBs at near
solar metallicity need to be produced in rather exotic binary
channels, or the magnetic effects are overestimated in our current
models.  The first option is not implausible, since reverse mass
transfer from the original secondary star onto the primary star during
its Wolf-Rayet phase \citep{2001A&A...369..939W}, or late stellar
merger may lead to an efficient core spin-up.  The realization
frequency of such events, however, even though it is difficult to
estimate, may still be small.  The latter would require a significant
angular momentum loss from the iron core, either during collapse or
from the proto-neutron star, in order to explain the relatively slow
rotation rates of young pulsars \citep[cf., however,][]{2004astro.ph..9422H}.

\acknowledgements

AH has been supported under the auspices of the U.S.\ Department of
Energy by its contract W-7405-ENG-36 to the Los Alamos National
Laboratory, by DOE SciDAC grant DE-FC02-01ER41176, and by NASA grant
SWIF03-0047-0037.


\bibliographystyle{aa}
\bibliography{2545gam}

\end{document}